\documentclass[prd,aps,twocolumn,preprintnumber]{revtex4}
\begin{document}
%\draft

\title{Seismic Search for Strange Quark Nuggets}
\author{Eugene T. Herrin}\affiliation {\textit{Geology Department, Southern Methodist University, Dallas, TX
75275}}
\author{ Doris C. Rosenbaum}\affiliation {\textit{Physics Department, Southern Methodist University, Dallas, TX 75275}}
\author{ Vigdor L. Teplitz}\affiliation {\textit{NASA Goddard Space Flight Center, Greenbelt, MD 20771}}
\date{Revised manuscript December, 2005}

\begin{abstract}
Bounds on masses and abundances of Strange Quark Nuggets (SQNs)
are inferred from a seismic search on Earth. Potential SQN bounds
from a possible seismic search on the Moon are reviewed and
compared with Earth capabilities. Bounds are derived from the data
taken by seismometers implanted on the Moon by the Apollo
astronauts. We show that the Apollo data implies
that the abundance of SQNs in the region of 10 kg to one ton must
be at least an order of magnitude less than would saturate the
dark matter in the solar neighborhood. 
\smallskip


\end{abstract}
\maketitle

\section{Introduction}

It has now been more than two decades since 1984 when Witten
raised the question of the existence of Strange Quark Matter (SQM)
as the possible ground state of baryonic matter \cite{witten}.  In
the interim, searches have been made in accelerators, stars and
other exotic and non-exotic locales.  They have been unsuccessful
in discovering evidence for SQM as well as in the harder task of
demonstrating its non-existence.  This paper addresses limits on
SQM that have been, and that might be, established by seismology
-- on Earth and on the Moon \cite{sig}.

SQM might be bound at zero pressure.  Nuclear matter made of up,
down and strange quarks would have the same potential energy from
the color force as nuclear matter made from just up and down, but
would have three Fermi seas instead of two.  With just up and
down, nuclear matter is not bound, but rather condenses into
protons and neutrons which, in turn, form finite sized nuclei at
zero external pressure.  With up, down and strange, on the other
hand, it might well be that, at zero pressure, there is binding of
large assemblies of quark matter.  SQM binding is also aided by
the fact that it tends to be electrically neutral except for
effects from the fact that the strange quark mass is larger than
those of the up and down.  The nuclear physics of SQM was worked
out in 1984 by Farhi and Jaffe \cite{fandj} and elaborated by
others since; see, for example \cite{weber} for a recent review
and references.  The argument for SQM binding has recently become
stronger from the realization that the color force should be
expected to form ``color-flavor locked" Cooper pairs thereby
increasing the binding over that expected from the earlier work
\cite{frank},\cite{rapp}; see Alford \cite{alf} for review and
references.

De Rujula and Glashow \cite{drg} outlined, in 1984, a variety of
places in which one might search for nuggets of SQM (SQNs),
including accelerators \cite{sandw}, mica \cite{price}, and among
cosmic rays \cite{mad4}. For these, results have so far been all
negative.  The cosmic ray search would be significantly augmented
were the AMS spectrometer \cite{sandw} deployed in orbit.  Reference \cite{drg} also pointed out that SQN passage
through the Earth, or other body, would cause a seismic signal
which, for large enough SQN mass, would be detectable. This paper
addresses that phenomenology.

The plan of the paper is as follows: in the remainder of this
section we review our past work on seismic detection of SQNs
passing through the Earth which was based on seismic reports
collected by the U.S. Geological Survey over the years 1990-1994.
In Section II, we address the question of what limits can be
placed on the distribution of SQNs from that work.  Given the
currency of the President's Exploration Initiative (see, for
example, \cite{mission}), in Section III we move on to the Moon,
reviewing our work in \cite{betal} on the relative sensitivities
of Earth and Moon for detecting SQNs and also presenting new
limits on SQN abundance deduced from the data collected by the
five seismometers implanted by the Apollo astronauts \cite{ls},
\cite{nasa}.  Section IV gives a brief summary.

We turn now to our past work on SQNs. In 1996, two of us used a
Monte Carlo calculation \cite{handt} to investigate the
sensitivity of the Earth in detecting SQN passage seismically.  We
took Cherenkov radiation as the model for generating seismic
waves since the galactic virial velocity, about $250$ km/s, is
about 25 times the speed of sound (and seismic waves) in the
Earth.  Since it takes six variables to specify an SQN passage
(time of entry, entry point, direction, and speed), we asked that
seven or more seismic stations detect the passage and that each
subset of six stations determine the same chord for the passage.
Note from Fig. 1 in \cite{handt} the difference in order of first
arrival times between point (epicentral) events and epilinear
ones.  We considered essentially all real stations at that time.
We approximated their sensitivities into three classes with
capabilities to detect $0.133, 1.33,$ and $13.3$ erg/s-cm$^2$
respectively.  We generated 120,000 random sets of values for the
6 parameters cited.  We found that 97 percent of detections were
from class 1 stations (the most sensitive ones).  We asked for the
lowest mass that would yield detection by 7 or more stations.  The
results are tabulated in Table I of the present work and also partially 
displayed graphically in Figure 1. Roughly, we
found that ten percent of the minimum masses, $m_{min}$, were below one metric
ton, 30 percent were below ten and 90 were below 220.  These results are sensitive to the assumption made for the fraction $\epsilon$ of nugget energy loss converted to detectable seismic waves ($m_{min}\sim\epsilon^{-3/2}$)

The U.S. Geological Survey kindly made available seismic reports
received from around the world from 1981 through 1993.  We
investigated in some detail the last 4 years of that data
\cite{andetal}.  These consisted of roughly a million reports,
including first signal arrival times, that had not been associated into Earthquakes (and about twice that number that had).  One of
our collaborators calculated travel time tables for signals
originating from deeper tham $750$ km (above that they have been
tabulated for some time); see \cite{tib}.  We eliminated reports
within an hour of large Earthquakes.  We tried all candidate
lines, in meshes of increasing fineness near good fits, seeking
ones that would minimize a figure of merit consisting of the sum
of the squared differences between reported times of first signal
arrival and calculated (on the basis of the assumed chord) times
of first signal arrival.  We eliminated all chords such that the
wave travel path from the point of closest approach would involve
passage through the Earth's iron core where travel times are not
so well understood.

The result of Reference \cite{andetal} was one candidate event
with a very good fit and waveforms that seemed to add additional
evidence of SQN passage.  The four most sensitive (and world-class)
stations in Australia
recorded strong signals from the event in question.  The arrival times
did not fit the spherical wave expected from a point source.  They did,
however, fit that expected from a line source.  It was later 
discovered, however, after publication
of Reference \cite{andetal}, that one of the four historically 
reliable stations had a large clock error (offset) for the entire 
month in which the
candidate set of reports occurred \cite{selbyetal}.  After
correcting for that offset, there was a good fit to an Earthquake.
That is, when that station was deleted, the remaining three 
Australian stations, as well as stations in South America, had
arrival times that did fit a point source.
The final result, therefore, was that no SQN passages were
detected in 4 years of seismic data.  In Section II below, we use
this fact combined with
the results of the Monte Carlo calculation in Table I to determine
limits on SQN abundance in the region of the galaxy near the sun.
\section{Limit on Number Reaching the Earth}

We estimate the bound on the number of strange quark nuggets
(SQNs) in our region of the galaxy implied by the negative results of Anderson
\textit{et al.} \cite{andetal}. That bound will naturally depend
on the nugget mass distribution.  If that distribution is skewed
too much toward low mass nuggets there will be a shortage of
nuggets capable of leaving detectable seismic signals and the
abundance bound will be weak.  If, on the other hand, it extends
too far toward large masses only a relatively small number will be
needed to saturate the abundance needed for galactic dark matter
DM and the bound from the absence of seismic events will again be
weak.  This point is made in \cite{handt}.

To estimate the bound on SQNs, we need convolve an assumed
abundance function $n(m)$, the number of nuggets of mass $m$ per
unit volume, with a probability $p(m)$ for detecting a nugget of
mass $m$ incident from a random direction.  We have for the number of nuggets that should be detected in time T

\begin{eqnarray}
 \label{npn}
N = 4\pi R_E^2(v/4)T\int{p(m)[dn(m)/dm]dm}
\end{eqnarray}

\noindent where $R_E$ is the radius of the Earth, $v$ is the
galactic virial velocity, and here $T$ is the four year period
over which Anderson {\itshape et al.} searched unassociated
seismic reports. We assume nuggets are distributed between $m_{min}$ and $M_{MAX}$ with $dn/da=Ka^{-\gamma}$ where $a$ is nugget
radius.  We take the normalization constant $K$ in terms of the
local density of dark matter, $\rho_{DM}$, approximately
$5\times10^{-25}$g cm$^{-3}$. The factor of $1/4$ takes account of
the fraction of SQNs per unit volume that will hit the nearby
Earth. We work in the approximation of all nuggets having the
galactic virial velocity (about $250$ km/s).
\begin{eqnarray}
\label{dndm}
dn/dm = Km^{-(\gamma+2)/3}
\end{eqnarray}
%\end{equation} with
\begin{eqnarray} \label{keq}
K=[(4-\gamma)/3]\rho_{DM}/[M_{MAX}^{(4-\gamma)/3} -m_{min}^{(4-\gamma)/3}]
\end{eqnarray}

We also need to specify the probability of seismic detection
$p(m)$ as a function of nugget mass $m$. We do this by means of
the Monte Carlo results of \cite{handt}.  These are given in
Table I and Figure 1.  With them, we can evaluate Eq. (\ref{npn}).  The
results are given in Tables II, III and IV for a few values of $\gamma $ around
4.  Note that $\gamma$ is the exponent in the distribution in terms of nugget radius.  The
value 3.5 is special, as found by Dohnanyi \cite{dohn}: if that
is the distribution given by particle collisions it will be
maintained under continuing collisions between the collision
fragments. The value $\gamma=4.0$ is also special in that the
integral for K becomes a logarithm.

\begin {table}
\caption{\label{TI} Distribution of minimum detectable masses   for 120,000 random events.}

\begin{ruledtabular}

\begin{tabular}{cccc} Mass & Number of events & Fraction &  Cumulative fraction\\
\hline

      0.063&         9&     0.000&     0.000\\
      0.100&        26&     0.000&     0.000\\
      0.158&        69&     0.001&     0.001\\
      0.251&       200&     0.002&     0.003\\
      0.398&       513&     0.004&     0.007\\
      0.631&      1043&     0.009&     0.016\\
      1.000&      2031&     0.017&     0.033\\
      1.585&      3449&     0.029&     0.061\\
      2.512&      5345&     0.045&     0.106\\
      3.981&      7462&     0.062&     0.169\\
      6.310&      9281&     0.078&     0.246\\
     10.000&     10678&     0.089&     0.335\\
     15.849&     11612&     0.097&     0.433\\
     25.119&     11939&     0.100&     0.532\\
     39.811&     11613&     0.097&     0.630\\

     63.096&     10180&     0.085&     0.715\\
    100.000&      8747&     0.073&     0.788\\
    158.489&      7027&     0.059&     0.847\\
    251.189&      5459&     0.046&     0.892\\
    398.107&      3988&     0.033&     0.926\\
    630.958&      2879&     0.024&     0.950\\
   1000.001&      2101&     0.018&     0.967\\
   1584.894&      1486&     0.012&     0.980\\
   2511.888&      1006&     0.008&     0.988\\
   3981.075&       659&     0.006&     0.994\\
   6309.580&       475&     0.004&     0.998\\
  10000.011&       267&     0.002&     1.000\\
\end{tabular}
\end{ruledtabular}
\end{table}

In Tables II, III and IV, we give the results for $\gamma =3.0,
4.0, $ and $5.0$. The individual tables use the $\gamma$ values
just cited.  In each, the rows and columns correspond to $m_{min},
M_{MAX}$ values in the limit of the integral in Eq. (\ref{npn}).
That is, SQNs are distributed in mass with index ${\gamma}\prime
=(\gamma+2)/3$ from $m_{min}$to $M_{MAX}$, not confined to just
the mass values of Table I, i.e. to values such that there is
chance of seismic detection on Earth.  In the tables, the top row
has $m_{min}=10^{-1.2}$ tons (corresponding to the lowest value in
Table I) and each succeeding row below has $m_{min}$ in Eq.(
\ref{keq}) down by a factor $10$. Similarly, column 1 has
$M_{MAX}=10^{4}$ tons, the highest value in Table I, and each
 succeeding column has $M_{MAX}$ up by a factor of $10$. Tables II-IV 
are made under the assumption that
the fraction $\epsilon = 0.05$ of SQN energy loss is turned into
seismic waves. Mass values in the tables should be multiplied by
(0.05/$\epsilon)^{3/2}$ for other assumptions [see
Eq.(\ref{Pm})below].

Tables II-IV indicate that, over the four year period studied by
\cite{andetal}, a value of $\gamma$ near $4$ would have produced
detectable nuggets for a fairly wide range in $M_{MAX}$ and
$m_{min}$.  For other $\gamma$ not too far from $\gamma=4$,
significant areas of the $[M_{MAX},m_{min}]$ plane should
similarly have produced detectable nuggets. A summary statement of
the results recorded in Tables II-IV is that Reference
\cite{andetal} precludes distributions (of a total density $\rho_{DM}$ of SQNs) with $\gamma$ in a small
range about $4.0$ and places some restrictions on distributions
with $\gamma$ near that range.

\begin {table}

\caption{\label{TII} Number of events expected for $\gamma=3$ as
$M_{MAX},m_{min}$ vary.  The 1-1 element has upper
and lower masses $M_{MAX},m_{min}$, in the distribution equal to
the first and last masses in Table I.  Succeeding rows (columns)
decrease (increase) $m_{min}(M_{MAX})$ by $10$.}
\begin{ruledtabular}

\begin{tabular}{cccccccccc}

   3.0&  1.4&  0.6&  0.3&  0.1&  0.1&  0.0&  0.0&  0.0&  0.0\\
   3.0&  1.4&  0.6&  0.3&  0.1&  0.1&  0.0&  0.0&  0.0&  0.0\\
   3.0&  1.4&  0.6&  0.3&  0.1&  0.1&  0.0&  0.0&  0.0&  0.0\\
   3.0&  1.4&  0.6&  0.3&  0.1&  0.1&  0.0&  0.0&  0.0&  0.0\\
   2.9&  1.4&  0.6&  0.3&  0.1&  0.1&  0.0&  0.0&  0.0&  0.0\\
   2.9&  1.4&  0.6&  0.3&  0.1&  0.1&  0.0&  0.0&  0.0&  0.0\\
   2.9&  1.4&  0.6&  0.3&  0.1&  0.1&  0.0&  0.0&  0.0&  0.0\\
   2.9&  1.4&  0.6&  0.3&  0.1&  0.1&  0.0&  0.0&  0.0&  0.0\\
   2.9&  1.4&  0.6&  0.3&  0.1&  0.1&  0.0&  0.0&  0.0&  0.0\\
   2.9&  1.4&  0.6&  0.3&  0.1&  0.1&  0.0&  0.0&  0.0&  0.0\\

\end{tabular}
\end{ruledtabular}
\end{table}

\begin {table}

\caption{\label{TIII} Number of events expected for $\gamma=4$ as
$M_{MAX},m_{min}$ vary. The 1-1 element has upper
and lower masses $M_{MAX},m_{min}$, in the distribution equal to
the first and last masses in Table I.  Succeeding rows (columns)
decrease (increase) $m_{min}( M_{MAX})$ by $10$.}
\begin{ruledtabular}

\begin{tabular}{cccccccccc}

  10.6&  8.7&  7.3&  6.3&  5.5&  4.8&  4.3&  3.9&  3.5&  3.2\\
   9.0&  7.6&  6.5&  5.7&  5.0&  4.5&  4.0&  3.6&  3.3&  3.0\\
   7.9&  6.8&  5.9&  5.2&  4.7&  4.2&  3.8&  3.4&  3.1&  2.9\\
   7.1&  6.2&  5.4&  4.8&  4.4&  3.9&  3.6&  3.3&  3.0&  2.8\\
   6.4&  5.7&  5.1&  4.5&  4.1&  3.7&  3.4&  3.1&  2.9&  2.7\\
   5.9&  5.3&  4.7&  4.3&  3.9&  3.5&  3.2&  3.0&  2.8&  2.5\\
   5.5&  4.9&  4.4&  4.0&  3.7&  3.4&  3.1&  2.9&  2.6&  2.5\\
   5.2&  4.6&  4.2&  3.8&  3.5&  3.2&  3.0&  2.8&  2.6&  2.4\\
   4.9&  4.4&  4.0&  3.7&  3.4&  3.1&  2.9&  2.7&  2.5&  2.3\\
   4.6&  4.2&  3.8&  3.5&  3.2&  3.0&  2.8&  2.6&  2.4&  2.2\\

\end{tabular}
\end{ruledtabular}
\end{table}

\begin {table}
\caption{\label{TIV} Number of events expected for $\gamma=5$ as
$M_{MAX},m_{min}$ vary. The 1-1 element has upper
and lower masses $M_{MAX},m_{min}$, in the distribution equal to
the first and last masses in Table I.  Succeeding rows (columns)
decrease (increase) $m_{min}( M_{MAX})$ by $10$.}
\begin{ruledtabular}
\begin{tabular}{cccccccccc}

  15.8& 15.6& 15.6& 15.5& 15.5& 15.5& 15.5& 15.5& 15.5& 15.5\\
   7.3&  7.3&  7.3&  7.3&  7.2&  7.2&  7.2&  7.2&  7.2&  7.2\\
   3.4&  3.4&  3.4&  3.4&  3.4&  3.4&  3.4&  3.4&  3.4&  3.4\\
   1.6&  1.6&  1.6&  1.6&  1.6&  1.6&  1.6&  1.6&  1.6&  1.6\\
   0.7&  0.7&  0.7&  0.7&  0.7&  0.7&  0.7&  0.7&  0.7&  0.7\\
   0.3&  0.3&  0.3&  0.3&  0.3&  0.3&  0.3&  0.3&  0.3&  0.3\\
   0.2&  0.2&  0.2&  0.2&  0.2&  0.2&  0.2&  0.2&  0.2&  0.2\\
   0.1&  0.1&  0.1&  0.1&  0.1&  0.1&  0.1&  0.1&  0.1&  0.1\\
   0.0&  0.0&  0.0&  0.0&  0.0&  0.0&  0.0&  0.0&  0.0&  0.0\\
   0.0&  0.0&  0.0&  0.0&  0.0&  0.0&  0.0&  0.0&  0.0&  0.0\\

\end{tabular}
\end{ruledtabular}

\end{table}

A second, in some sense opposite, way of presenting the results of
\cite{andetal} is that of de Rujula and Glashow \cite{drg}.  They
take all SQN of one mass – with an abundance that yields the dark
matter density in the solar neighborhood (approximately
$5\times10^{-25}$gm cm$^{-3}$).  This is a useful tool for
comparisons even though, in real life, we would expect a
distribution in mass. It could correspond, in some approximation
to primordial SQN dark matter production for which one might
expect mass determined by the number of quarks within the horizon
at the time of production.

We will make comparisons with seismic
detection on the Moon in the next section.  If $\rho_{DM}$ is all
in SQNs of mass $m$, and if $p(m)$ is a theta function, we have, in 4
years, from Eq. (\ref {npn})

\begin{eqnarray}
 \label{Nhit}
dN_{hits}/dt&=&(\rho_{DM}/m)(v/4)4\pi R_E^2 p(m)\nonumber\\& \rightarrow&
(5\times10^8/m)/yr
\end{eqnarray}

\noindent where $m$ is in grams. The limit in Eq. (\ref{Nhit}) is
for large $m$ for which the probability of detection goes to one.
Using the cumulative fraction in the results of the Monte Carlo of
\cite{handt} in Table I gives the results of Figure 2.

Figure 2, conservatively,
 shows that Earth should be a reasonable detector of
 DM SQNs if they are peaked in mass about a value in the range 
$0.15$ to $150$ tons. For smaller
mass, it is only for special geometries that the seismic signals
are detected, while for larger mass the abundance of incident SQNs
is too small. Finally, it is important to emphasize that we have
assumed in our results that the fraction $\epsilon$ of SQN energy
loss in Earth passage that goes into seismic signals is $0.05$. It
could be considerably higher as discussed in Reference
\cite{andetal} thereby decreasing the minimum detectable mass by
$(\epsilon/0.05)^{-3/2}$.  It could also, of course, be lower.  A 
reliable calculation of $\epsilon$ would appear to be an important
goal.

We believe that a continuation of this search effort with
terrestrial seismology should make use of real time seismic data
now available from most seismic stations, rather than the old data
used by \cite{andetal}.  However, as will be discussed in the
following two sections, a better approach to seismology might be
to apply it to other solar system bodies with lower seismic
backgrounds and hence the capability to detect nuggets of smaller
mass. These nuggets are likely to be more abundant in any
distribution, if SQM is indeed bound at zero pressure.

\section{Seismic Detection on the Moon and Beyond}

The Apollo astronauts implanted five seismometers at various
locations on (the near side of) the Moon.  These functioned for
several years and give some picture of lunar seismic activity.  In
brief, there are weak, deep quakes caused by the tides as the
Moon's position relative to the Earth and Sun varies; there are
(relatively strong and infrequent) shallow quakes caused by
unknown geologic processes (it is believed there is no tectonic
activity), and there are impacts.  There is no background from
winds and waves.  This feature means that seismometers used on
Earth should be sensitive to seismic waves of \textit{amplitude}
about one third as great if used on the Moon. In this section, we
first review the implications of this fact for the seismic search
for strange quark nuggets drawing on our discussion in Banerdt
\textit{et al.} \cite{betal}.  A second important lunar seismic
feature is the Moon's tendency to ``ring" for some time after
seismic excitation. We leave this feature for later study.
Seismology on the Moon is reviewed on the NASA Johnson Space
Flight Center web site, in the Apollo summary of the Moon
\cite{nasa}, as well as in standard text books such as Carroll and
Ostlie \cite{carost}.

 In this section we focus primarily on
the implications of the Apollo seismic bounds, numbers of seismic
events (around 2500/yr) and total lunar yearly seismic energy. The
measured amount of the latter, or, more precisely, the amount
directly inferred from the measurements made, is $10^{17}$ ergs
per year, which can be compared to $2\times10^{24}$ ergs for the
Earth.  It should be noted, however, that, as pointed out by
Nakamura \cite{naka}, the actual figure could be, on the average,
several orders larger if one extrapolates the curve of numbers of
relatively strong, shallow quakes as a function of shallow quake
magnitude.  For present purposes, however, we just address the
question as to the extent to which the observed limit gives
information on the abundance of SQNs in our part of the Galaxy. We
note that this question was raised in discussion at the Caltech
Jet Propulsion Laboratory April, 2004, Physics in Solar System
Exploration  conference held in Solvang, California.
 We begin by briefly reviewing, from Reference \cite{betal} the major factors
that enter in the relative sensitivities of the Earth and the Moon
when used as seismic SQN detectors.  These include:

\noindent 1. Relative cross sectional areas of Earth and Moon;\\ 
2. Likely numbers of seismic stations and station \\placement;\\ 
3. Earthquake backgrounds;\\ 
4. Ocean and atmospheric backgrounds;\\ 
5. Attenuation with distance;\\ 
6. Effective blackout of signals by Earth's iron core; and\\ 
7. Lunar ringing.\\

\noindent We address each of these items in turn below.

1. Areas. The ratio of the cross sectional area of Moon to Earth
is $\alpha_{M}/\alpha_{E} \sim 0.075$.

2. Numbers and placement of stations.  Assume about 10 seismic
stations for good coverage of the Moon. The number of stations
needs to be considered both per unit area (Moon wins if stations
can be affordably placed optimally since Earth has no sensitive
stations in or near oceans) and in the context of 7 or more
station reports needed both to fix and to confirm the 6 nugget
trajectory parameters (Earth wins). We take the rough
approximation that these two factors cancel. We believe that this
approximation is conservative in the sense that it likely favors
the Earth and penalizes the Moon.

3. Earthquake backgrounds. Anderson \textit{et al.} \cite{andetal}
found it desirable to remove all station reports within one hour
of a quake of magnitude 4.0 or more because of the difficulty of
reliably identifying reverberations. The result was to remove
signals from 1/3 of the minutes in the year. Low seismic activity
frees the Moon of such a penalty but see item 7 below.

4. Ocean and atmospheric background. This very important factor
means that seismic detection on the Moon is only limited by
instrument noise and ringing (below). The relative contributions
of atmospheric noise and instrument noise (ocean noise is less
than atmospheric inland) is unknown. We estimate that atmospheric
(amplitude) noise is the greater effect by about an order of
magnitude in energy.

5. Attenuation with distance. Since seismic energy falls, as with
other forms of energy, with distance as $r^{-2}$, seismic
amplitude falls as $1/r$ making Earth seismic signals received at
a station weaker, on average, than those received at lunar
stations by the ratio of the radii ($0.273$).

6. Iron core blackout. We compute for the Earth the volume of the
cone segment $z \sim [r_E-r_C/2, 2r_E]$ (where the ratio of the
core radius, $r_C$ to the Earth radius $r_E$ is about $0.5$) from
which seismic signals will not reach a station at $z = 0$ in
reliably predictable times (because we have found no reliable way
of following propagation of SQN signals through the Earth's iron
core). The result, weighting with the attenuation factor, is that
about one third of signals are eliminated for the Earth. There is
controversy with regard to a possible, relatively small iron core
for the Moon. We assume/approximate that none is present.

7. Lunar ringing. Seismic signals on the Moon exhibit codas. These
persisted for some time with Apollo instruments.  We do not know
the rate of decrease of these signals for small values.  The codas
could set a lower limit on achievable sensitivities. Additionally,
at greater sensitivities they could require a subtraction
procedure as in item 3 above. We do not attempt here to quantify
these issues.

We include these considerations as needed below.  Our aim is to be
relatively conservative in our estimates of lunar capability for
seismic SQN detection.  We assume that seismic signals on the Moon
can be detected to a factor of $\sqrt{10}$ in amplitude of ground
motion below those on Earth. The factor of $\sqrt{10}$ in
amplitude implies a factor of 10 below in energy, making modern
seismometers on the Moon sensitive to signals on the order of
0.013 erg/cm$^2$\,s.  Note that significant further sensitivity
improvement would be possible with, for example, superconducting
technology. In addition to the sensitivity improvement, for
identifying epilinear signal generation we assume only minimal
seismometer emplacements, say 6 or 7 widely separated.  Thus we
require signal strength sufficient to be detected at distances of
$2R_M$, the lunar diameter.  As above, we set $\epsilon$, the
fraction of SQN energy loss converted to detectable seismic waves,
at $\epsilon\sim0.05$ except as noted.

With these assumptions, the minimum detectable mass $m_d$ for all
transit trajectiories  for an SQN with galactic virial speed $v_V$
can be found from equating its signal strength to our assumed
instrument noise

%\begin{equation}
\begin{eqnarray}
\label{Pm} P_m &=& 1.3\times10^{-2} erg\,cm^{-2}s^{-1}\nonumber\\
    &=& \epsilon\rho_M\pi
 (3m_d/4\pi\rho_N)^{2/3}
 v_V^3/4\pi R_M^2 \sim 10^{-6}m_d^{2/3}
\end{eqnarray}
%\end{equation}
 This gives (with nugget density $\rho_N = 2\times
10^{14} g/cm^3$)
\begin{eqnarray}
\label{md} m_d \sim 125 kg
\end{eqnarray}
%\end{equation} where $m_d$ scales as $\epsilon ^{3/2}$. So that
$\epsilon = 0.1$ would imply $m_d \sim 50$ kg and $\epsilon = 0.5$
3 kg. Nuggets of mass below $m_d$ might also be detected depending
on the location of their transit trajectories. For m $>$ $m_d$,
detection probability would be 100\%.

 We assume below that, for nuggets of mass $m_d$ the
detection probability on the Moon is one for $m>m_d$ and zero for $m<m_d$.  This is consistent with
our requirement for signal detection at distances $2R_M$.  More
detailed modeling and more nuanced discussion would need mass and velocity distributions.  We
consider first the lunar companion to Figure 2, the number of SQN
detections, for given single SQN mass $m$ and for sufficient
abundance to constitute the local DM density.

%\begin{equation}
\begin{eqnarray}
\label{dndtm} dN/dt = (\rho_{DM}/m)(v/4)4\pi R_M^2yr \sim
(3\times10^7/m)/yr
\end{eqnarray}
%\end{equation}

At the lower mass sensitivity limit, we would expect about 600
events each year if $\rho_{DM}$ were in the form of $50$ kg SQNs
with $\epsilon = 0.1$. The mass, $m_d$ decreases as $ P_m^{3/2}$
so another factor of 10 or so decrease would bring the minimum
detectable signal down to kilograms. 
Equations (\ref{md}, \ref{dndtm}) say that the Moon and the Earth
are somewhat complementary as SQN detectors. The Earth with its
larger area gets eight times the events for the same abundance
assumption while the Moon has the sensitivity to detect
significantly lower nugget mass.  Together they span a range of
roughly $10^4$ at least in mass detection.

We turn now to the ``Apollo limits,'' limits on SQNs that can be
inferred from the Apollo results that:

    ---As noted above, about 2500 seismic events per year were detected in the
three categories: deep, weak, tidal Moonquakes; shallow, relatively strong Moonquakes; and impacts.  We assume that a population of tens of SQNs over 50kg or even somewhat less would have been identified as an additional class of seismic events.

    ---The total lunar seismic energy in a year inferred from the data taken was about $2\times10^{17}$erg (compared with $10^{24}$ erg for Earth).

We consider here only these two gross characteristics, ignoring
more subtle arguments that might be exploited.  The first, the
limit on numbers of events in an unrecognized class, implies
roughly that the abundance of SQNs with masses in the range
$10-10^3$ kg must be at least an order of magnitude less than
would be required to saturate the local DM density. However there
were too few seismometers to be certain that some of the events
identified as deep quakes were not actually SQN passages.

Moving to the second Apollo limit, we have that the seismic energy from each lunar SQN passage is, on the average, given by

%\begin{equation}
\begin{eqnarray}
E_S = \epsilon(v_V^2/2)\rho_MR_M\pi(3m/4\pi\rho_N)^{2/3} \sim 5\times10^{12}m^{2/3}
\end{eqnarray}
%\end{equation} 

where we have assumed an average SQN travel
distance in the Moon of $R_M$.  Putting this together with Eq.
(\ref{dndtm}), we see that the seismic energy in one year, from
SQNs of mass $m$ would be given by

%\begin{equaiton}
\begin{eqnarray}
\label{etlim} E_T = (3\times 10^7/m)(5\times10^{12}m^{2/3}) \sim
10^{20}/m^{1/3}
\end{eqnarray}
%\end{equation}

Equation (\ref{etlim}) implies that the abundance of SQNs in the
range 10 kg to a ton must be at least an order of magnitude less
than that that would saturate the local DM density.  It is a much
stronger limit than that from numbers of events, since the latter
depends on identifying a class of SQN events from about ten sets
of seismic reports, while the former just says that their
contribution to total seismic energy would be noticed. These points are given graphically
in Figure 3 where events expected if all local DM is in mass m nuggets
are compared with the Apollo limits on maximum number of nuggets
found from the bound on total amount of seismic energy in a year.
Note the desirability of determining instrument noise reliably and on 
decreasing it so as to enlarge the mass region in which DM can be bounded.
Recall from Section II we expect that, if SQNs are dark matter, they should be
relatively narrowly peaked in mass distribution around the mass of
baryons $M_B$ within the horizon at formation time and temperature.
$M_B$ is given by $M_B \sim 2\times 10^{21}/T(GeV)^3$ (see Appendix A in
\cite{kt}). Thus, in principle, the range of SQM dark matter formation
temperatures covered by the Apollo data is roughly $10^5$ - $10^6$
GeV and by the Earth data $10^4$ - $10^5$ GeV. These temperature
values would vary as $(\epsilon /0.05)^{3/2}$ for $\epsilon \neq
0.05$.  

However, it is important to note that current understanding of quantum chromodynamics implies that the strong coupling constant decreases with increasing energy so that we would not expect nugget formation for temperatures over a few hundred MeV by which time the horizon contains about $10^{24}$ grams of baryons.  This would imply dark matter of nuggets with masses about one percent the mass of the Earth and sizes around a meter.  Thus the range of SQN masses that can be investigated by terrestrial and lunar seismology does not, in all likelihood, include that appropriate for SQM as DM.  Nor, of course do any of the other methods now available such as those discussed in [8].  All of course do, however, include detecting fragments from colliding neutron stars if these are made from SQM, an eventuality considered more likely than SQM DM.  In spite of the fact that DM is unlikely to be SQNs in the mass ranges to which the Earth and moon are sensitive, it is convenient to express limits on SQM abundance in terms of $\rho_{DM}$

Could, however, $10^{24}$ gm DM ever be detected (in the spirit of exploring the moon, the solar system and beyond)?  Equation (7) implies that, if the number density of DM particles is given by $\rho_{DM}/m$, the mass of DM that can be detected (say 10 events per year) by a system of size $R$ is

\begin{equation}
	m \sim 10^7\rho_{DM}v_VR^2 \sim 10^{-10}R^2
\end{equation} 

Helioseismology ($R_{\odot}\sim10^{11}$cm) could reach to a million metric tons providing nuggets of the corresponding size (millimeters) would leave behind a sufficiently energetic, detectable, identifiable signal.  To reach to $10^{24}$ grams would appear to call for raising $R$ by a factor of $10^6$, i.e. to $10^4$ astronomical units (the size of the inner radius of the Oort cloud) – not a near term project.

\section{Summary}

We briefly recapitulate our results.

--- The work of Anderson \textit{et al.} \cite{andetal} precludes SQN
distributions in our neck of the galaxy given by Eq. (\ref{dndm})
with $ \gamma$ in a small range about 4 and places some
restrictions for $\gamma$ near that range, e.g. 3 and 5.

--- Anderson \textit{et al.} \cite{andetal} should have found 5 or more epilinear events for $10^5$ gm $< m < 3\times 10^8$ gm  if all the local dark matter density were in SQNs in that range and 0.05 of SQN energy loss is
into seismic waves.

--- As pointed out in Ref. \cite{betal}, deployment of
seismometers on the Moon with instrument noise in the region
$10^{-2}$ erg/cm$^2-s$ would mean certain detection of SQNs down
to the 125 kg level while ones with $10^{-4}$erg/cm$^2-s$ would
permit detection to the 125 gm level providing persistence of
lunar codas (ringing)did not create too great a background.

--- The total yearly lunar seismic
energy being $10^7$ less than that of Earth implies that the
abundance of SQNs in the range 10 kg to a ton must be at least an
order of magnitude less than would saturate the dark matter
density in the solar neighborhood.

\begin{acknowledgments}

We very much appreciate discussions with D. Anderson, W. Banerdt,
T. Chui, H. Paik, K. Penanen, J. Sandweiss, and D. Strayer.  The
work of Section III was greatly aided by J. Ormes and N. White
suggesting we review helioseismology and lunar seismic data
generally, by the participant at the JPL Solvang Conference
who raised the issue of the bound on total lunar seismic energy, and by helpful suggestions from the anonymous Phys. Rev. referee.

\end{acknowledgments}

FIG. 1: Curves give the fraction and cumulative fraction of SQN's of mass $m$ SQNs that would have been detected by terrestrial seismic stations existing in early 1990s.

FIG. 2: Curve uses Eq. (9) with probability of detection $p(m)$ from the Monte Carlo results (Table I and Fig. 1) to give SQN detections that should have been seen, in 1990-1993, if local DM were mass $m$ SQNs.

FIG. 3 Solid curve: (the log of) the number of lunar SQN passages expected if local DM were mass $m$ SQNs. Dashed curve: number of mass $m$ passages permitted by the Apollo bound on total lunar seismic energy. For $m>1000$, no limit is implied. 

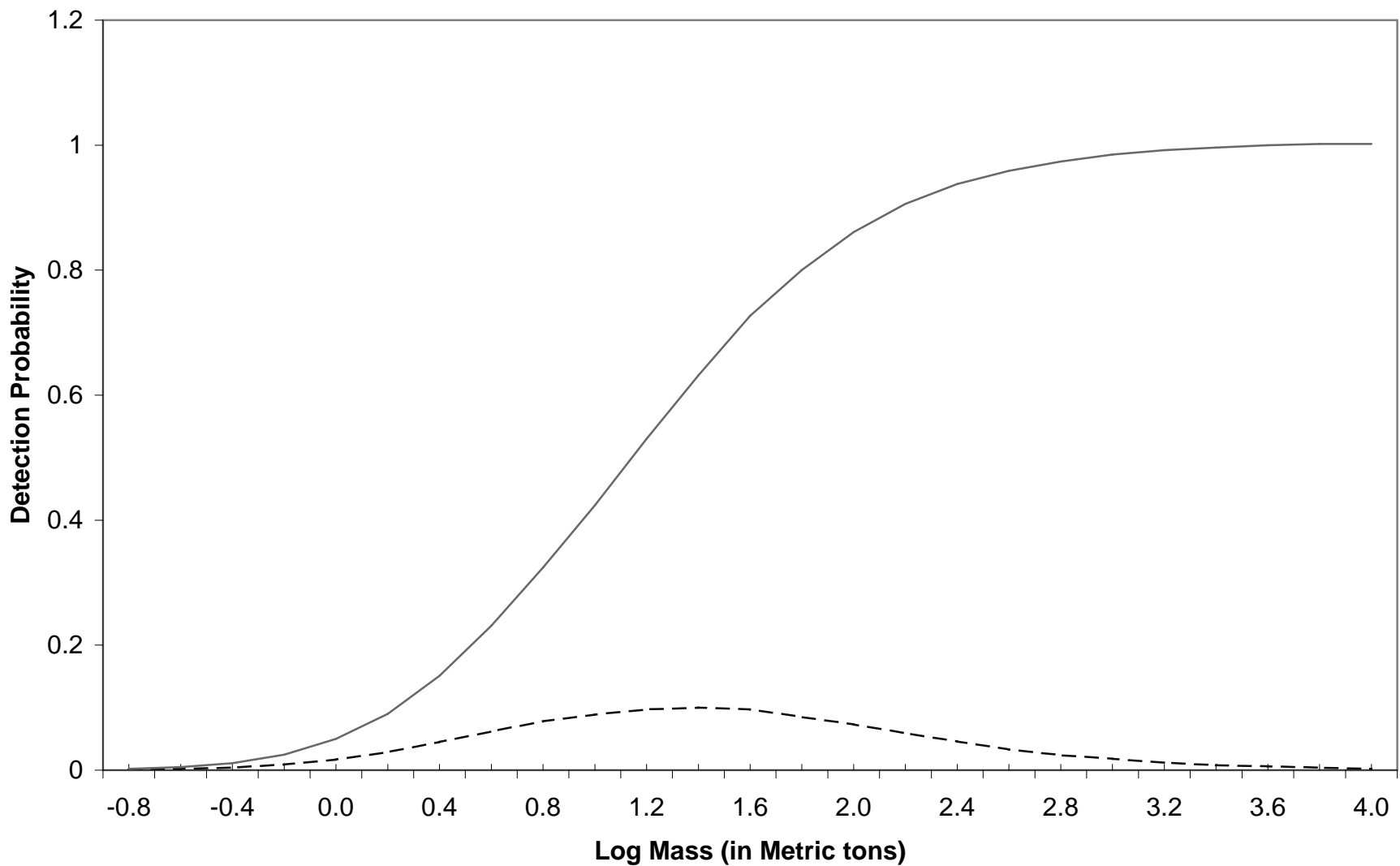
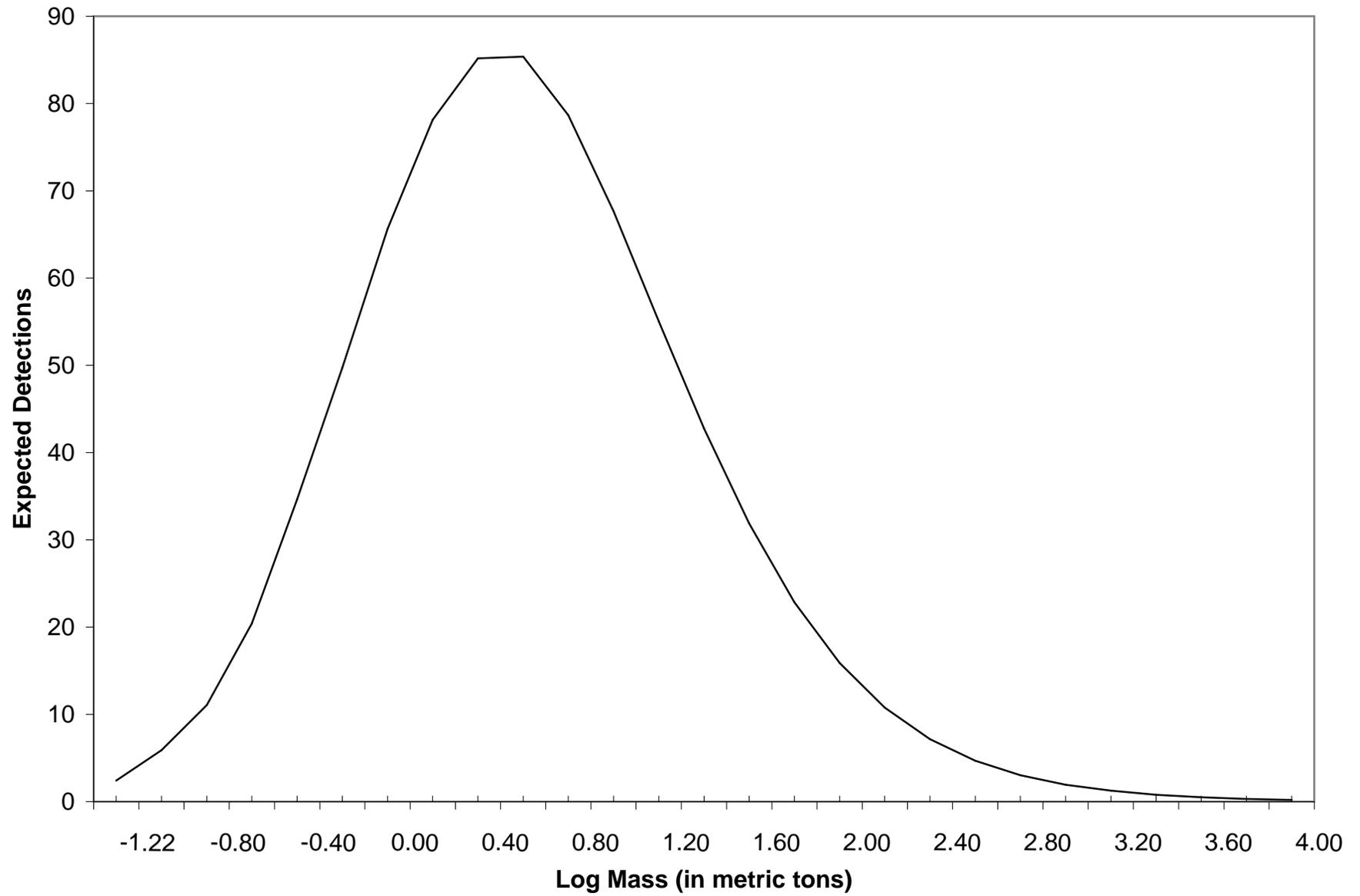
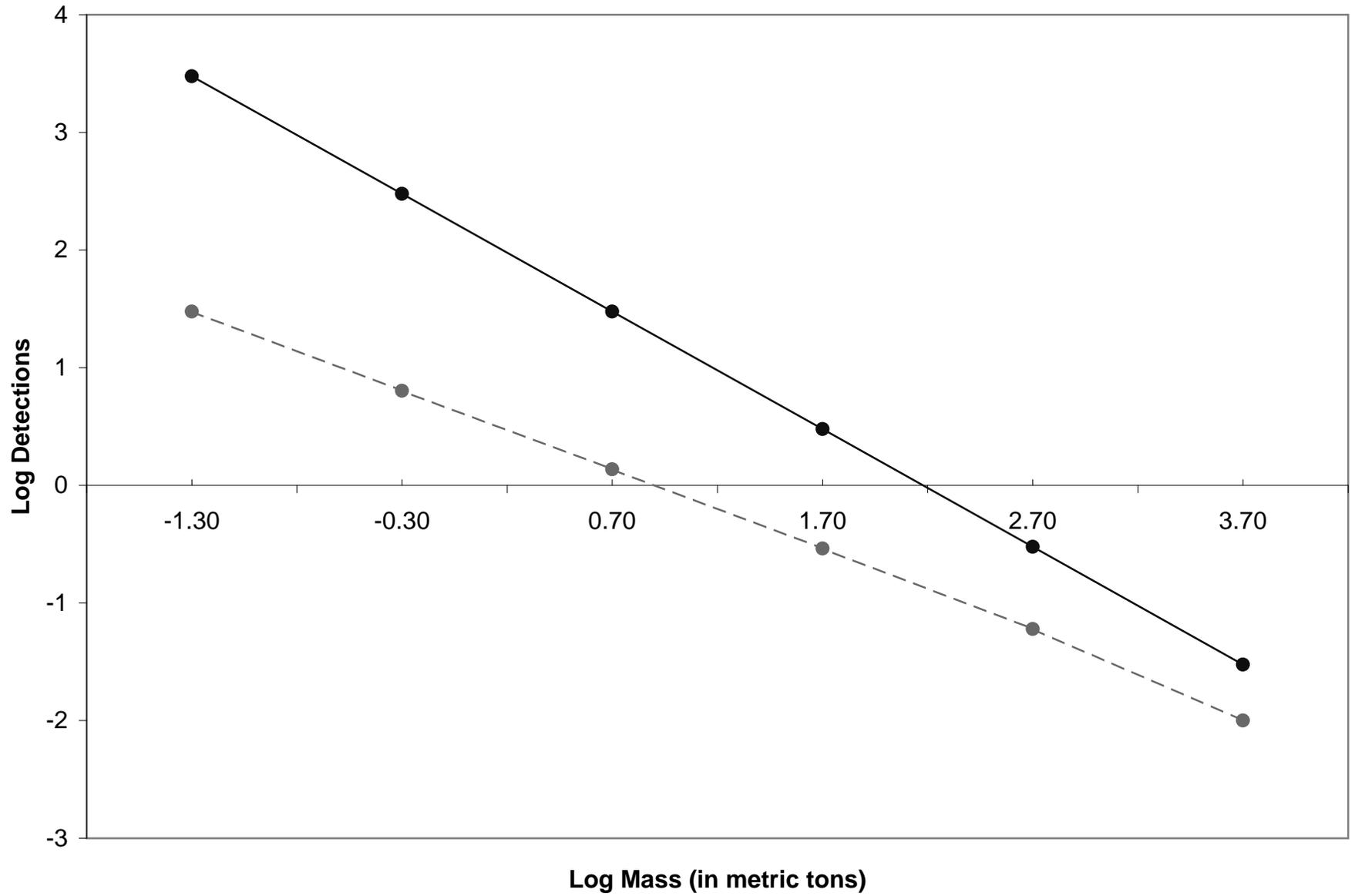
\end{document}